\documentclass[a4paper,11pt]{article}
\usepackage[a4paper,left=2.73cm,right=2.7cm,top=3cm,bottom=3.5cm]{geometry}

\usepackage{amsfonts,amsmath,amssymb,tabstackengine,amsthm}
\usepackage{array}
\usepackage{comment}
\usepackage{mathrsfs}
\usepackage[bb=boondox]{mathalfa}
\usepackage{cite}

\usepackage{MnSymbol}
\usepackage{tensor}

\usepackage[colorlinks=true,linktocpage=true,linkcolor=blue,citecolor=blue]{hyperref}
\usepackage{graphicx}

\addtocontents{toc}{\protect\setcounter{tocdepth}{2}}
\numberwithin{equation}{section}

\theoremstyle{definition}

\def\Es{{\mathrm{E}_{7(7)}}}
\def\es{{\mathfrak{e}_{7(7)}}}

\def\SL{{\mathrm{SL}}}
\def\sl{{\mathfrak{sl}}}

\def\SU{{\mathrm{SU}}}
\def\su{{\mathfrak{su}}}
\def\SO{{\mathrm{SO}}}
\def\so{{\mathfrak{so}}}
\def\GL{{\mathrm{GL}}}
\def\Mint{{M_{\text{int}}}}
\def\Hprin{H_{\text{prin}}}

\newcommand{\scalP}[2]{{\langle #1 ,\,#2 \rangle}}

 \usepackage [latin1]{inputenc}

\begin{document}
\begin{titlepage}

\thispagestyle{empty}

\begin{center}
{\LARGE \textbf{de Sitter and S-folds in type IIB\\ from $\mathcal{N}=4$ $D=4$ gauged supergravity}}

\vspace{40pt}
		
{\large \bf  {\large \bf Colin Sterckx}}
		
\vspace{25pt}

{\normalsize
Universit\'e Libre de Bruxelles (ULB) and International Solvay Institutes,\\
Service  de Physique Th\'eorique et Math\'ematique, \\
Campus de la Plaine, CP 231, B-1050, Brussels, Belgium.}
\\[7mm]

\texttt{colin.sterckx@ulb.be}

\vspace{40pt}

\abstract{
\noindent In this paper, we classify the type IIB uplifts of the $\mathcal{N}=4$
$D=4$ pure $\SO(4)$-gauged supergravity admitting a dS$_4$ solution. As for their
AdS cousins, the internal geometries of these uplifts consist of two $S^2$ fibred
over a Riemann surface $\Sigma$, and are classified by pairs of harmonic functions
$h_{1,\,2}$ on $\Sigma$, providing a new family of de Sitter solutions in type IIB.
Choosing specific functions $h_{1,\,2}$ on $\Sigma = \mathbb{R} \times I$, we
perform an S-fold quotient of the internal space along its non-compact direction.
This S-folding allows us to evade the assumptions of the Maldacena-Nunez no-go
theorem and provides a classical de Sitter solution of type IIB supergravity with
finite and non-vanishing four-dimensional effective Newton constant.}

\end{center}
\end{titlepage}

\tableofcontents

\hrulefill
\vspace{10pt}

\section{Introduction}

Finding stable scale-separated de Sitter vacua is one of the most important missing links between string theory and cosmological observations. Despite large classes of supersymmetric and non-supersymmetric AdS solutions being known, de Sitter solutions are bound by strong constraints even at the level of two-derivative supergravity. In particular, assuming a geometric, compact and sufficiently regular internal space, the Maldacena-Nunez no-go theorem precludes the existence of de Sitter solutions of type II supergravities in the absence of additional ingredients\cite{Maldacena:2000mw}.

In this paper, we explore a new way to avoid this obstruction based on a non-geometric quotient of the internal space. We construct ``S-folded" de Sitter solutions of type IIB supergravity, i.e. solutions such that a direction in the internal space can be quotiented up to an $\SL(2,\,\mathbb{Z})$ duality transformation. These S-folds have played an important role in the construction of AdS backgrounds \cite{Inverso:2016eet,Guarino:2019oct,Bobev:2019jbi,Guarino:2020gfe,Bobev:2020fon,Arav:2021tpk,Guarino:2021kyp,Giambrone:2021zvp,Arav:2021gra,Bobev:2021yya,Guarino:2021hrc,Cesaro:2021tna,Bobev:2021rtg,Cesaro:2022mbu,Bobev:2023bxs,Guarino:2024gke} and have proven to be an essential ingredient for their holographic realisation\cite{Assel:2018vtq}. Moreover, these S-folds have also been used to challenge other Swampland conjectures such as the non-SUSY AdS conjecture \cite{Ooguri:2016pdq,Giambrone:2021wsm}, showing them to be an interesting playground to test our ideas about quantum gravity.

Our starting point is four-dimensional pure $\mathcal{N}=4$ gauged supergravity. This theory contains six vector fields in the gravity multiplet and its duality group is 
\begin{equation}
    G_D =  \SL(2)\times \SO(6) \,.
\end{equation}
A subgroup of this duality group can be gauged and, using the embedding tensor formalism, it has been shown to admit exactly three inequivalent types of $\SO(4)$ gaugings \cite{Inverso:2015viq}. These families of gaugings are a refinement of the ``de Roo-Wagemans" angle \cite{deRoo:1985jh}. The first family of gaugings admits an anti-de Sitter vacuum \cite{Louis:2014gxa} and was first uplifted to type IIB in \cite{Guarino:2024gke} (with a more general classification of its uplifts performed in \cite{Rovere:2025jks}). The second family is a family of Freedman-Schwarz-like gaugings \cite{Freedman:1978ra} which are parametrised by the ratio of two $\SO(3)$ coupling constants. That family does not admit maximally symmetric solutions due to a runaway scalar potential. Finally, the last type of gauging admits a de Sitter solution. In this paper, we will classify the uplifts of this theory in type IIB.

The technical framework is that of consistent truncation in exceptional generalised geometry (EGG) \cite{Cassani:2019vcl,Coimbra:2011nw,Coimbra:2012af,Hohm:2013vpa,Hohm:2013uia,Hohm:2014fxa}. To perform the classification of the uplifts, we will follow the methods presented in \cite{Rovere:2025jks}. This method yields, via the uplift of the dS$_4$ solution, solutions to the type IIB equations of motion on
\begin{equation}
    \text{dS}_4 \times \Sigma \times S^2 \times S^2\,,
\end{equation}
with a warping depending on the $\Sigma$ coordinates as well as non-trivial fluxes and a non-trivial axio-dilaton profile. Exactly as for the AdS$_4$ solutions of \cite{DHoker:2007hhe,DHoker:2007zhm}, the uplifts presented here, and their corresponding dS$_4$ solutions, will depend on two harmonic functions $h_1$ and $h_2$ on $\Sigma$, providing us with a new class of dS$_4$ solutions in type IIB.

We then address the de Sitter no-go theorem by choosing specific functions $h_1$ and $h_2$ on 
\begin{equation}
    \Sigma = \mathbb{R} \times I\,.
\end{equation}
The resulting geometric construction possesses a non-compact direction as required by the no-go theorem. However, it also enjoys a discrete symmetry generated by a shift along this direction combined with an $\SL(2,\,\mathbb{Z})$ transformation. Performing the quotient with respect to this symmetry, a process known as ``S-folding", we obtain a solution with finite internal volume and a finite and non-vanishing effective four-dimensional Newton constant. Neither higher-derivative corrections nor non-perturbative ingredient are thus needed to avoid the no-go.  

Nevertheless, there are important limitations concerning this solution. First, the family of dS$_4$ solutions we present is not scale-separated. Second, the curvature of the de Sitter S-fold becomes large as one reaches the boundaries of $\Sigma$. Hence, the supergravity approximation cannot be trusted arbitrarily close to these regions. Third, since the solution is non-supersymmetric (as are all de Sitter backgrounds), its perturbative stability is not guaranteed. Indeed, from the four-dimensional description one finds at least two scalar modes with negative mass squared, rendering the solution perturbatively unstable \cite{Roest:2009tt}. It follows that our solution should be understood as a classical construction demonstrating the possibility of finite-$G_N^{(4)}$ non-geometric de Sitter backgrounds, and not as a controlled string theory vacuum.

This paper is organised as follows. In section 2, we will review the pure $\mathcal{N}=4$ $D=4$ gauged supergravities and the classification of their $\SO(4)$ gaugings. In section 3, we will perform the classification of the type IIB uplifts of the theory with the de Sitter solution. We will provide detailed formulas for the ansatz in Exceptional Field Theory, comparing the de Sitter and the anti-de Sitter gaugings. In section 4, we will provide the full uplift formulas in plain supergravity language. In section 5, we will consider a specific dS$_4$ uplift and show that S-folding that solution renders its internal manifold compact. We will study several quantities such as the effective Newton constant, the internal volume and internal proper lengths, as well as the regime of validity of the supergravity approximation. We conclude in section 6 with a discussion of the results and possible directions for future work. Appendix A collects our conventions for $\Es$ generators.

\section{Pure \texorpdfstring{$\mathcal{N}=4$ $\SO(4)$}{N=4 SO(4)} gauged supergravities in four dimensions}

We summarise the bosonic sector of $\mathcal{N}=4$ $D=4$ pure gauged supergravity (i.e. without vector multiplets) \cite{Schon:2006kz,DallAgata:2023ahj} with a focus on its different SO(4) gaugings. The global symmetry group of this supergravity is 
\begin{equation}
    G_D =  \SL(2)_D \times \SO(6)\,.
\end{equation}
The bosonic sector contains a metric, $g$, six electric vector fields and their six dual magnetic vectors, transforming in the representation $R_v=(\mathbf{2},\,\mathbf{6})$  of $G_D$ and denoted $A_\mu^{A}$, as well as 2-forms in the adjoint representation of $G_D$. Finally, the theory contains a complex scalar field $\tau = - \chi + i\,e^{-\xi}$, parametrising the coset space $\SL(2)_D/\SO(2)$. In what follows, the index $_I=1,\,\dots,\,6$ labels the vector representation $(\mathbf{6})$ of $\SO(6)$ while the index $_a = {}_{\pm}$ labels the fundamental representation $(\mathbf{2})$ of $\SL(2)_D$, i.e. $_A = _{aI}$. The generators of $G_D$ in the representation $R_v$ will be denoted $(t_\alpha)_A{}^B$.

\subsection{Classification of \texorpdfstring{$\SO(4)$}{SO(4)} gaugings}

We start by recalling the classification of $\SO(4) \subset \SO(6)$ gaugings of this theory (see e.g. \cite{Inverso:2015viq}). The gauging is given by a linear map
\begin{equation}
    \theta: R_v \rightarrow \mathfrak{g}_D\,,
\end{equation}
called the \emph{embedding tensor}. This map appears in the definition of the covariant derivative:
\begin{equation}
    D = d + A^A \theta_A\,.
\end{equation}
Introducing indices, this map is a tensor $\theta_{a I}{}^\alpha$ (equivalently $X_{AB}{}^C = \theta_{A}{}^\alpha (t_\alpha)_B{}^C$) satisfying a set of linear and quadratic constraints. Since we are not gauging a subgroup of $\SL(2)_D$, the embedding tensor constraints reduce to
\begin{equation}
    \begin{split}
        \theta_{a I}{}^\alpha (t_\alpha)_{JK} = \theta_{a [I}{}^\alpha (t_\alpha)_{JK]} =: f_{a [IJK]} \hspace{5mm}&\Leftrightarrow \hspace{5mm}\theta \in (\mathbf{2},\,\mathbf{10}_+) \oplus(\mathbf{2},\,\mathbf{10}_-) \,\footnotemark,\\
        [\theta_A,\,\theta_B] = -X_{AB}{}^C \theta_C \hspace{5mm}&\Leftrightarrow\hspace{5mm} f_{a I[JK} f_{b| LM]}{}^{I} = 0 = \epsilon^{ab} f_{aIJK} f_{bLM}{}^I\,.
    \end{split}
\end{equation}
\footnotetext{The rank-three antisymmetric representation of $\so(6)$ is reducible and splits into a self-dual and an anti-self-dual piece denoted $(\mathbf{10}_{\pm})$.}
\noindent Moreover, the image of $\theta$ must be isomorphic to a subalgebra $\so(4)\subset \so(6)_D$. There are two inequivalent embeddings of $\so(4)$ in $\so(6)_D$, however, only one contains singlets in the embedding tensor representations. Since the quadratic constraints imply that the embedding tensor must be invariant under the gauge group transformations, only one of the embeddings can be consistently gauged while preserving supersymmetry. The relevant branchings are
\begin{equation}
\begin{array}{ccc}
\SL(2)_D \times \SO(6)& \rightarrow &\SL(2)_D \times \SO(3) \times \SO(3) \\
    (\mathbf{2},\,\mathbf{6}) &\rightarrow &(\mathbf{2},\,\mathbf{1},\,\mathbf{3}) \oplus (\mathbf{2},\,\mathbf{3},\,\mathbf{1})\,,\\
    (\mathbf{2},\mathbf{10}_{\pm}) &\rightarrow &(\mathbf{2},\,\mathbf{3},\,\mathbf{3}) \oplus  (\mathbf{2},\,\mathbf{1},\,\mathbf{1}) \,.
\end{array}
\end{equation}
Thus, there are four singlets under $\SO(4)_g (\cong \SO(3) \times \SO(3))$ which can be turned on in the embedding tensor. To write down the four-parameter family of gaugings, we split $I \rightarrow i \oplus \underline{i}$ where $i$ and $\underline{i}$ label the fundamental representation of each $\SO(3)$ factor respectively. The non-zero entries in the embedding tensor read:
\begin{equation}
\label{eq:embTensorN4}
    f_{a\, ijk} = g_a \epsilon_{ijk}\hspace{1cm} f_{a\, \underline{ijk}} = \underline{g}{}_a \epsilon_{\underline{ijk}}\,.
\end{equation}
The quadratic and linear constraints are satisfied for any choice of real constants $g_a$ and $\underline{g}{}_a$. The scalar potential reads\footnote{We use the standard convention $\epsilon_{+-} = \epsilon^{+-} = 1$.}
\begin{equation}
    V = -\frac{1}{2} \left(M^{ab}(g_a g_b + \underline{g}{}_a \underline{g}{}_b) + 4 \epsilon^{ab} g_a \underline{g}{}_b\right)\,,
\end{equation}
where
\begin{equation}
    M_{ab} = e^{\xi}\begin{pmatrix}
        e^{-2\xi}+\chi^2 & -\chi\\
        -\chi & 1
    \end{pmatrix}\,.
\end{equation}
Taking into account the global $\SL(2)_D$ dualities, and overall rescaling which can be reabsorbed by a trombone symmetry of the equations of motion, inequivalent gaugings are classified by
\begin{equation}
    (\mathbb{R}^2_0\times \underline{\mathbb{R}}^2_0)/(\SL(2)_D \times \mathbb{R}^\times) \,.
\end{equation}
This results in three inequivalent families of $\SO(4)_g$ gaugings:
\begin{itemize}
    \item AdS gauging: $g_+ = 1$ and $\underline{g}{}_- = 1$, with $g_- = 0 = \underline{g}{}_+$. This choice of gauging is of the type considered in \cite{Louis:2014gxa}. The scalar potential reads
    \begin{equation}
        V = -\frac{1}{2} \left(\text{Tr}(M_{ab})+4\right)\,.
    \end{equation}
    This theory admits an AdS vacuum at $\tau=i$ and has been uplifted to M-theory on $S^7$ in \cite{Cvetic:1999au} (although a full classification of M-theory uplifts is still missing). Its type IIB uplifts were fully classified in \cite{Rovere:2025jks} and correspond to consistent truncations around any of the solutions of \cite{DHoker:2007zhm,DHoker:2007hhe}.
    \item dS gauging: $g_+ = 1$ and $\underline{g}{}_-= -1$, with $g_- = 0 = \underline{g}{}_+$. The scalar potential reads
    \begin{equation}
        V = -\frac{1}{2} \left(\text{Tr}(M_{ab})-4\right)\,,
    \end{equation}
    The embedding tensor is related by an O$(6)$ transformation to the AdS gauging. Since the duality group is 
    $\SO(6)$, such a transformation does not leave the scalar potential invariant (even up to dualities). This is why this gauging admits a de Sitter solution at $\tau = i$. This gauging can be obtained as a consistent truncation of maximal $\mathcal{N}=8$ supergravity with $\SO(4,\,4)$ gauge group which originates from M-theory on a hyperboloid \cite{Hohm:2014qga,Baron:2014bya} (although a full classification of M-theory uplifts is still missing). The classification of its uplifts in type IIB will be presented in this paper. Finally, notice that the dS vacuum is a maximum of the scalar potential, implying that the corresponding uplifts are unstable (see \cite{Roest:2009tt}).
    \item Freedman-Schwarz gauging: $g_+ = \cos\theta$, $\underline{g}{}_+ = \sin\theta$ and $g_- = 0 = \underline{g}{}_-$ for $\theta \in (0,\,\pi/2)$ as described in \cite{Freedman:1978ra}. It arises as an $S^3\times S^3$ truncation of type IIA supergravity. The scalar potential reads
    \begin{equation}
       V= -\tfrac{1}{2} e^{\xi}\,.
    \end{equation}
    When $\theta = 0$ or $\tfrac{\pi}{2}$, the gauge group degenerates, reducing to an $\SO(3)$ gauging of the supergravity.
\end{itemize}

In this paper, we will focus on the classification of uplifts of the de Sitter gauging. As we will see, the structure of the end result will be reminiscent of the AdS case up to important changes of signs, allowing for dS$_4$ solutions. 

\subsection{The bosonic equations of motion}
The equations of motion for each of the three gaugings read
\begin{equation}
\begin{split}
    &\star \mathcal{H}^{aI} = - (\mathbb{C} \mathcal{M} \mathcal{H})^{aI}\,,\\
    &\mathcal{H}^{IJ} = 0\,,\\
    &\square \xi - e^{2\xi} \partial_\mu \chi \partial^\mu \chi - \partial_\xi V + \mathcal{H} \partial_\xi \mathcal{M} \mathcal{H} = 0\,,\\
    &\nabla_\mu\left(e^{2\xi} \partial^\mu \chi\right) - \partial_\xi V + \mathcal{H} \partial_\chi \mathcal{M} \mathcal{H} = 0\,.
\end{split}
\end{equation}
The two-forms $B^\alpha$ whose image is not in $\so(4)_g$ decouple from the theory. The twelve-dimensional scalar matrix $\mathcal{M}$ (not to be confused with $M_{ab}$) reads
\begin{equation}
    \mathcal{M} = -\begin{pmatrix}
        e^\xi & 0 & -\chi e^\xi& 0\\
        0 & e^\xi & 0 & - \chi e^{\xi}\\
        -\chi e^{\xi} & 0 & e^{\xi}|\tau|^2 & 0\\
        0 & -\chi e^\xi & 0 & e^{\xi}|\tau|^2
    \end{pmatrix} \otimes \mathbb{1}_{3\times 3}\,,
\end{equation}
while the symplectic matrix $\mathbb{C}$ is
\begin{equation}
    \mathbb{C} = \begin{pmatrix}
        0_{6\times 6} & \text{Id}_{6\times 6}\\
        -\text{Id}_{6\times 6} & 0_{6\times 6}
    \end{pmatrix}\,,
\end{equation}
and field strengths for the vectors are defined as 
\begin{equation}
\begin{split}    
    &\mathcal{H}^{ai} = dA^{ai} +\frac{1}{2} g_b \epsilon_{jk}{}^i A^{jb} A^{ka} + \frac{1}{2} g^a \epsilon_{ijk} B^{jk}\,, \\
    &\mathcal{H}^{a\underline{i}} = dA^{a\underline{i}} +\frac{1}{2} \underline{g}{}_b \epsilon_{\underline{jk}}{}^{\underline{i}} A^{\underline{j}b} A^{\underline{k}a} +\frac{1}{2}  \underline{g}{}^a \epsilon_{\underline{ijk}} B^{\underline{jk}}\,,
\end{split}
\end{equation}
with $g^a = g_b\epsilon^{ba} $ (and equivalently for $\underline{g}{}^a$). These equations have to be supplemented by the Einstein equation of motion.

\section{Classification of uplifts}

As explained in \cite{Cassani:2019vcl}, consistent truncations of type II and M-theory supergravities are better understood in the context of Exceptional Generalised Geometry (EGG) \cite{Coimbra:2011nw,Coimbra:2012af} and Exceptional Field Theory (ExFT)\cite{Hohm:2013vpa,Hohm:2013uia,Hohm:2014fxa}. They correspond to reductions of the structure group of the generalised tangent bundle with constant singlet intrinsic torsion. This intrinsic torsion is then identified with the embedding tensor of the reduced theory. In the case at hand, we will reduce the $\Es\times \mathbb{R}^+$ structure group down to $\SU(4)_S$ with a constant singlet intrinsic torsion identified with \eqref{eq:embTensorN4}. In this section, will follow the construction of \cite{Rovere:2025jks} (summarised in its section 3.4) to build the generalised sections corresponding the appropriate reduction of the structure group. The construction goes in three steps and the final formulas for the consistent truncation in type IIB are collected in the next section.

First, it has been shown that $\Mint$ must admit an $\SO(4)_g$ action. The action implies the existence of a principal stabiliser $\Hprin\subset \SO(4)_g$. For a compact gauge group, the choice of a principal stabiliser is unique for each type of uplift (in type IIB or M-theory). In this case, we will see that we must choose $\Hprin \cong \SO(2) \times \SO(2)$, meaning that the internal manifold will formed of two two-spheres fibred over a Riemann surface.

Second, using the previous choice of principal stabiliser, we will solve an algebraic condition for an element $E^\flat \in \Es\times\mathbb{R}^+$ which will be related to the generalised frame. This condition is
\begin{equation}
    \label{eq:algCondition}
    \mathbb{P} E^\flat \mathcal{E}^{\underline{m}} = \Theta^{\underline{m}}\,,
\end{equation}
where $\mathbb{P}$ is the projector of the $\mathbf{56}$ of $\Es$ on the $\SU(4)_S$ singlets and $\Theta^{\underline{m}}$ is the reduced embedding tensor. The $\underline{m}$ index labels a basis of $\mathfrak{so}(4)_g/\mathfrak{h}_{\text{prin}}$. 

Finally, one must solve a set of PDE on the quotient $\Sigma = \Mint/\SO(4)_g$ which read
\begin{equation}
    d(\mathbb{P} \cdot L \cdot E^\flat \cdot e^{-1})_{|(p)} = 0\,.
\end{equation}
where $L \in \Es$ is a coset representative of $\SO(4)_g/\Hprin$, $e^{-1}$ is its associated round vielbein and $E^\flat\in \Es\times \mathbb{R}^+$ is a function of $\Sigma$ solving \eqref{eq:algCondition}. The notation $V^M_{|(p)}$ stands for the $p$-form component of the generalised vector $V^M$. The solution to these equations will provide us with the appropriate generalised sections and the frame defining the consistent truncation. We refer to \cite{Rovere:2025jks} for the various assumptions, simplifications and methods which can be used to solve these algebraic and differential conditions.

We highlight a last technical point. There are various isomorphic representations of $\es$ (and $\Es$). Two of them will be useful for the discussion here and are referred to as the ``$\SL(8)$-basis" and the ``$\SU(8)$-basis". The $\SL(8)$-basis is useful when discussing the connection between generalised geometry and the standard geometry of supergravities in terms of metric and fluxes. The $\SU(8)$-basis is better suited to discuss the reduction of the structure group (because it is always possible to take $G_S \subset \SU(8)$). Our notations and the relation between the two bases are summarised in app. \ref{app:Eseven}.

\subsection{The structure group}

Generalised $\SU(4)_S$ structures were discussed in \cite{Malek:2017njj}. The $\SU(4)_S$ structure group embeds in $\Es$ as 
\begin{equation}
\label{eq:fundbranchN4IIB}
\begin{array}{ccccc}
E_{7(7)} &\rightarrow& \SO(6,\,6) \times \SL(2)_D &\rightarrow & \SU(4)_S \times \SU(4)_R \times \SL(2)_D\,,\\[2mm]
 \mathbf{56} &\rightarrow & (\mathbf{12},\,\mathbf{2}) \oplus (\mathbf{32'},\,\mathbf{1}) &\rightarrow&(\mathbf{6},\,\mathbf{1},\,\mathbf{2}) \oplus (\mathbf{1},\,\mathbf{6},\,\mathbf{2})  \oplus (\mathbf{4},\,\mathbf{\bar{4}},\,\mathbf{1}) \oplus (\bar{\mathbf{4}},\,\mathbf{4},\,\mathbf{1}) \,.
\end{array}
\end{equation}
In the $\su_8$ basis of $\es$, the $\su(4)_S$ generators act on the first four indices of the $\mathbf{8}_v$. For the projector on the $\SU(4)_S$ invariant space we use the standard normalisation
\begin{equation}
    \mathbb{P}_{aI}{}^M \mathbb{P}_{bJ}{}^N \Omega_{MN} = \delta_{IJ} \epsilon_{ab}\,.
\end{equation}
The matrix $\mathbb{P}$ projects onto the entries $v_{ab}$ and $v^{ab}$ of the $\mathbf{56}$ with $a,\,b=5,\,\dots,\,8$. The $\SL(2)_D$ group commuting with both $\SU(4)_R$ and $\SU(4)_S$ is generated by
\begin{equation}
    \sl(2)_D = \langle t^s_{[1234]},\,t^c_{[1234]},\,(t^v)_1{}^1 +\cdots + (t^v)_4{}^4 - (t^v)_5{}^5 - \cdots -(t^v)_8{}^8 \rangle\,.
\end{equation}
We embed the generators of the gauge group $\SO(4)_g \cong \SO(3)_1 \times \SO(3)_2$ in $\es$ as 
\begin{equation}
    \begin{split}
        &\tau_1 = t^{28}_{[56]}+t^{28}_{[78]},\hspace{5mm}\,\tau_2= t^{28}_{[57]}-t^{28}_{[68]},\,\hspace{5mm} \tau_3 = -t^{28}_{[58]}-t^{28}_{[67]}\\
        &\tau_{\underline{1}}=t^{28}_{[56]}-t^{28}_{[78]},\,\hspace{5mm} \tau_{\underline{2}}=t^{28}_{[57]}+t^{28}_{[68]},\,\hspace{5mm}\tau_{\underline{3}}=t^{28}_{[58]}-t^{28}_{[67]}
    \end{split}
\end{equation}
where $\mathfrak{so}(3)_1 = \langle\tau_i\rangle$ and $\mathfrak{so}(3)_2 = \langle\tau_{\underline{i}}\rangle$. With our conventions, these are the real generators in $\su(4)_R$ and they split into a self-dual and an anti-self-dual piece corresponding to each of the $\mathfrak{so}(3)$ factors.

As for the AdS gauging, the principal stabiliser is $\Hprin = \SO(2)_1 \times \SO(2)_2$ whose generators can be chosen to be $\tau_1$ and $\tau_{\underline{1}}$. Using standard angular variables, the coset representative we choose is
\begin{equation}
    L = \exp(-\phi_1 \,\tau_{1}) \cdot \exp((\theta_1 + \pi) \tau_{2})\cdot\exp(-\phi_2 \,\tau_{\underline{1}}) \cdot \exp((\theta_2 + \pi) \tau_{\underline{2}})\,,
\end{equation}
from which we read the standard vielbein one-forms on $S^2 \times S^2 \cong (\SO(3)_1\times \SO(3)_2)/(\SO(2)_1 \times \SO(2)_2)$:
\begin{equation}
    \mathring{e} = (d\theta_1,\,\sin\theta_1 d\phi_1 ,\, d\theta_2 ,\, \sin\theta_2 d\phi_2)\,.
\end{equation}
The internal manifold is of the form
\begin{equation}
    \Mint= S^2 \times S^2 \times \Sigma\,,
\end{equation}
where the two-spheres are warped over the Riemann surface $\Sigma$. 

Finally, using the standard angular variables $(\theta_i,\,\phi_i)$ on each $S^2_i$ it will prove useful to define the $\SO(3)_i$ equivariant functions
\begin{equation}
    \begin{array}{ll}
    Y^1 = \cos(\theta_1)\,,& Y^{\underline{1}} = \cos(\theta_2)\,,\\
    Y^2 =\sin(\theta_1)\, \sin(\phi_1) \,,& Y^{\underline{2}} = \sin(\theta_2)\,\sin(\phi_2)\,,\\
    Y^3 =\sin(\theta_1)\,\cos(\phi_1) \,,& Y^{\underline{3}} =\sin(\theta_2) \,\cos(\phi_2)\,.
    \end{array}
\end{equation}
corresponding to the embedding of each $S^2$ in $\mathbb{R}^3$. With these coordinates, the $\so(3)_1\oplus \so(3)_2$ Killing vectors read
\begin{equation}
\begin{array}{ll}
    k_1 = \partial_{\phi_1}\,, & k_{\underline{1}} = \partial_{\phi_2}\,,\\
    k_2 = \cos\phi_1\, \partial_{\theta_1} -\cot\theta_1\,\sin\phi_1 \partial_{\phi_1}\,,& k_{\underline{2}} =  \cos\phi_2\, \partial_{\theta_2} -\cot\theta_2\,\sin\phi_2 \partial_{\phi_2}\,, \\
    k_3 = -\sin\phi_1  \partial_{\theta_1} - \cot\theta_1 \cos \phi_1 \partial_{\theta_1}\,,\hspace{1cm} & k_{\underline{3}}=  -\sin\phi_2  \partial_{\theta_2} - \cot\theta_2 \cos \phi_2 \partial_{\theta_2}\,.
\end{array}
\end{equation}

\subsection{The ansatz}

In order to solve the algebraic condition \eqref{eq:algCondition}, we start by classifying compatible section constraints $\mathcal{E}_M{}^m$, i.e. solutions to the section constraints equations \cite{Hohm:2013uia}
\begin{equation}
    Y^{MN}{}_{PQ} \mathcal{E}_M \mathcal{E}_N = 0
\end{equation}
such that
\begin{equation}
    \mathbb{P}\mathcal{E}{}^{\underline{m}} = \Theta{}^{\underline{m}}\,.
\end{equation}
The non-vanishing entries of the reduced embedding tensor are
\begin{equation}
    \Theta_{1+}{}^1 = \Theta_{2+}{}^2 = 1 \hspace{5mm}\text{and}\hspace{5mm}\Theta_{\underline{1}-}{}^{\underline{1}} = \Theta_{\underline{2}-}{}^{\underline{2}} = -1\,.
\end{equation}
From there, it becomes clear that compatible section constraints for the de Sitter gauging can be obtained from the compatible section constraints for the anti-de Sitter gauging by acting on them with the $\GL(6)$ element reversing the orientation of the $\underline{1}$ and $\underline{2}$ direction.

\paragraph{Classifying compatible solutions to the section constraints}
We start from a standard solution to the section constraint $\mathcal{E}_{IIB}$ (provided in \cite{Inverso:2016eet} for example) which, in the $\SL(8)$ basis, has the following non-zero entries:
\begin{equation}
\begin{split}
&\mathcal{E}_{IIB\,[27]}{}^{1} = 1,\,\hspace{5mm}\mathcal{E}_{IIB\,[37]}{}^{2} = 1,\,\hspace{5mm}\mathcal{E}_{IIB\,[47]}{}^{3} = 1,\\
&\mathcal{E}_{IIB\,[57]}{}^{4} = 1,\,\hspace{5mm}\mathcal{E}_{IIB\,[67]}{}^{5} = 1,\,\hspace{5mm}\mathcal{E}_{IIB}{}^{\,[18]}{}^{1} = 1\,.
\end{split}
\end{equation}
One can obtain a compatible solution to the section constraint by using the section constraint
\begin{equation}
    \mathcal{E}_0 = E_0 \cdot \mathcal{E}_{IIB}\,,
\end{equation}
where
\begin{equation}
    E_0 = \sqrt{2}\,K_0 \cdot (S_{\su \rightarrow \mathfrak{sl}})^\dagger \cdot  \,\text{exp}\left[\tfrac{1}{2}\log 2\, (t_6{}^6 + t_{7}{}^7 - t_{1}{}^1 - t_8{}^8)\right] \cdot s\,,
\end{equation}
\begin{equation}
\label{eq:defE0}
\begin{split}
    K_0 =& \text{diag}(i^{-1/2},\,i^{-1/2},\,i^{1/2},\,i^{1/2},\,i^{1/2},\,i^{1/2},\,i^{-1/2},\,i^{-1/2})\\
   & \cdot R_{[12]}\left(\tfrac{-\pi}{4}\right)\cdot R_{[34]}\left(\tfrac{\pi}{4}\right) \cdot R_{[56]}\left(\tfrac{-\pi}{4}\right)\cdot R_{[78]}\left(\tfrac{\pi}{4}\right)\cdot R_{[24]}\left(\tfrac{\pi}{2}\right)\\
   &\cdot R_{[68]}\left(\tfrac{-\pi}{2}\right)\cdot R_{[15]}\left(\tfrac{-3\pi}{8}\right)\cdot R_{[26]}\left(\tfrac{-3\pi}{8}\right)\cdot R_{[37]}\left(\tfrac{-\pi}{8}\right)\cdot R_{[48]}\left(\tfrac{-\pi}{8}\right)\,.
\end{split}
\end{equation}
and 
\begin{equation}
    s = \text{diag}(1,1,-1,-1,1,1)\in \GL(6)_{IIB} \subset \Es\,.
\end{equation}
We denoted by $R_{[ab]}(\alpha)$ the $\SO(8)\subset \SU(8)$ rotation of angle $\alpha$ in the plane $[ab]$. The $\SU(8)$ matrix $K_0$ could be further simplified by multiplying it by an irrelevant $\SU(4)_S$ element from the left. However, this would complicate later computations hence the version we present here. The change of basis matrix $S_{\su \rightarrow \mathfrak{sl}}$ is defined in App. \ref{app:Eseven}.

From the compatible section constraint $\mathcal{E}_0$, we obtain an embedding of $\Hprin$ in $H_{\mathcal{E}_0} \subset \GL(6)_{\mathcal{E}_0}$. This group is generated by
\begin{equation}
    \tilde{\tau}_1 = t^{28}_{[12]} +t^{28}_{[34]}+t^{28}_{[56]}+t^{28}_{[78]} \hspace{5mm}\text{and}\hspace{5mm}\tilde{\tau}_{\underline{1}}=t^{28}_{[12]}-t^{28}_{[34]}+t^{28}_{[56]}-t^{28}_{[78]}\,.
\end{equation}
The next step is to classify $\text{Stab}(\mathcal{E}^{H}_\text{comp})$, the stabiliser of the space of $H$-invariant compatible section constraint. As in the AdS case, this space can be characterised as
\begin{equation}
    \mathcal{E}^H_{\text{comp}} = \SL(2)_Z \times \GL(2)_\Sigma
\end{equation}
where $\GL(2)_\Sigma \subset \GL(6)$ acts on the $\Sigma$ directions of the embedding tensor and we compute that the $\SL(2)_Z$ group is generated by the elements
\begin{equation}
\label{eq:sl2Zgen}
   \sl(2)_Z = \langle z_0 = t^s_{[1278]},\,z_1=t^v_{11} +t^v_{22} +t^v_{77} +t^v_{88} -t^v_{33} -t^v_{44} -t^v_{55} -t^v_{66},\,z_2= t^c_{[1278]} \rangle \,.
\end{equation}
Acting on $\mathcal{E}_0$, this group generates the full set of $H$-invariant compatible solution to the section constraints. We will denote by 
\begin{equation}
    Z: \SL(2)_Z \rightarrow \SL(2)_Z\subset  \Es
\end{equation}
the map embedding $\SL(2)_Z$ in $\Es$.\footnote{By abuse of notation, we will use the same notation for the map $Z : \sl(2)_Z \rightarrow \sl(2)_Z \subset\es$ which should not cause confusion.} In this basis, we denote the generators of $\sl(2)_Z$ by
\begin{equation}
    z_0 = Z\begin{pmatrix}
        1 &0 \\
        0& -1
    \end{pmatrix}\,,\hspace{5mm}z_1 = Z\begin{pmatrix}
        0 &1 \\
        -1& 0
    \end{pmatrix}\,,\hspace{5mm}z_2 = Z\begin{pmatrix}
        0 & 1\\
        1 & 0
    \end{pmatrix}\,.
\end{equation}

\paragraph{Fluxes}

Now, we can characterise the generalised $H_{\mathcal{E}_0}$-invariant fluxes, i.e. the elements of $\Es\times \mathbb{R}^+$ leaving $\mathcal{E}_{IIB}$ invariant. Expressed in the $\SL(8)$ basis, these generators can be ordered with respect to their weight under $\GL(6)$ rescaling and correspond to type IIB fluxes. These fluxes, upon performing some gauge fixing, are specified by the group element 
\begin{equation}
    \mathcal{S} = \exp\left[\rho^{-2}\left(b_{11} t_{[1236]} + b_{12} t_{[1456]} + b_{21} t_{[8236]} + b_{22} t_{[8456]}\right)\right] \cdot \exp(\rho^{-4}\,\lambda\, t_6{}^7) 
\end{equation}
where $b_{ij}$, $\lambda$ and $\rho$ are functions of $\Sigma$ only. The factors of $\rho$ are related to the trombone symmetry and this definition will prove convenient in what comes next. Finally, one should parametrise the $\SL(2)_{IIB}$ elements corresponding to the type IIB axio-dilaton. The IIB axio-dilaton coset representative $\mathcal{V}_{\SL(2)_{IIB}}$ is generated by the $\sl_8$ generators $t_1{}^8,\,t_{8}{}^1$ and $t_1{}^1 - t_{8}{}^8$\,.

\paragraph{The ansatz}
The final ansatz for the generalised frame reads
\begin{equation}
\label{eq:AnsatzFrameFinal}
     E =\rho\hspace{1mm} L \cdot Z\cdot E_0 \cdot \mathcal{V}_{SL(2)_{IIB}} \cdot \mathcal{S}  \cdot e^{-1}\,,
\end{equation}
where $Z$ is a general element of $\SL(2)_Z$ and
\begin{equation}
    e = \text{diag}\left(\rho\,,\rho\sin\theta_1,\,s\,\rho,\,s\,\rho\sin\theta_2,\,\rho^{-1} \,l,\,\rho^{-1}\,l\right)_{\GL(6)_{IIB}}\,.
\end{equation}
Finally $\rho$ and $l$ are simply a positive real functions of $\Sigma$. Notice that we have used coordinate transformations to fix the metric on $\Sigma$ to be conformal. Fixing $s=1$ corresponds to the AdS gauging while $s=-1$ corresponds to the dS gauging and we see that the dS ansatz (not its solution) simply differs from the AdS ansatz by a $\GL(6)_{IIB}$ transformation (equivalently, this corresponds to including or not the element $s$ from the definition of $E_0$ \eqref{eq:defE0}). This explains why the principal stabilisers $\Hprin$ in the dS and AdS cases are isomorphic. In what follows we will leave $s$ free, only imposing $s^2=1$, in order to compare the type IIB uplifts of the AdS and the dS gaugings.

\subsection{The differential constraint}
Plugging the ansatz \eqref{eq:AnsatzFrameFinal} in the differential constraint
\begin{equation}
    d(\mathbb{P} \cdot E)_{|(p)} = 0 \,,
\end{equation}
we obtain a series of PDE on $\Sigma$ for the fluxes, axio-dilaton, $\rho$, $\lambda$ and $Z$. Up to gauge transformations or global $\SL(2,\,\mathbb{R})$ symmetries, the solutions to these differential constraints can be expressed by providing a pair of harmonic functions on $\Sigma$: $h_1$ and $h_2$. We introduce two real coordinates $(\alpha, \,\eta)$ on $\Sigma$. We also introduce the harmonic dual of these functions $h_1^D$ and $h_2^D$. These functions are uniquely defined up to an additive constant via the identity
\begin{equation}
    dh_1 = - \star dh_1^D \hspace{5mm}\text{and}\hspace{5mm}dh_2 = - \star dh_2^D\,.
\end{equation}
Equivalently, they are the two functions such that the functions $\mathcal{A}_{1,\,2}$
\begin{equation}
    2\mathcal{A}_1(z) = h_1^D + i\,h_1\hspace{5mm}\text{and}\hspace{5mm}2\mathcal{A}_2(z) = h_2 - i h_2^D\,
\end{equation}
 are holomorphic with respect to $z = \eta + i\, \alpha$.

The solution to the torsion constraint reads
\begin{equation}
\begin{split}
   & \mathcal{V}_{SL(2)_{IIB}} = \begin{pmatrix}
      0 & \sqrt{h_1/h_2}\\
      -\sqrt{h_2/h_1} & 0
   \end{pmatrix}\,, \\[2mm]
   & Z = 
       \frac{1}{\sqrt{- s \star_\Sigma(d\log h_1 \wedge \star_\Sigma d\log h_2)}} \begin{pmatrix}
           -s \partial_\eta \log h_2 & \partial_\alpha \log h_1\\
           s \partial_\alpha \log h_2 & \partial_\eta \log h_1
       \end{pmatrix}
   \,,\\[4mm]
    & \rho = (h_1\,h_2)^{1/4}\,,\hspace{5mm} l = 8 (- s \star_\Sigma(d h_1 \wedge \star_\Sigma d h_2))^{1/2}\,,\\[2mm]
    &b_{12} = 0 = b_{21},\,\hspace{5mm}b_{11} = \sqrt{2} \,s\,h_1^D \,,\hspace{5mm} b_{22} = \sqrt{2}\, s\, h_2^D \\[2mm]
    &d\lambda=\,-\begin{pmatrix}
        h_1^D\\h_2^D
    \end{pmatrix} \epsilon \begin{pmatrix}
       d h_1^D\\d h_2^D
    \end{pmatrix}+\begin{pmatrix}
        h_1\\h_2
    \end{pmatrix} \epsilon \begin{pmatrix}
        dh_1\\dh_2
    \end{pmatrix}\,,\\[2mm]
    & s = \begin{cases}1 &\text{ for the AdS gauging}\\
    -1 &\text{ for the dS gauging}\end{cases}\,.
\end{split}
\end{equation}
One can always solve (locally) for $\lambda$ because the r.h.s. is closed and only $d\lambda$ will appear in gauge invariant quantities. We immediately observe that we obtain well-defined solutions only if 
\begin{equation}
\label{eq:posDefCondition}
    h_1 h_2 > 0 \hspace{5mm}\text{and}\hspace{5mm} - s \star_\Sigma (dh_1 \wedge \star_\Sigma dh_2) > 0\,.
\end{equation}

\subsection{The invariant sections}

At first glance, it would appear that we could solve the torsion constraint by using the sections corresponding to the AdS gauging and performing an O(6) transformation on them sending $K_{\pm\underline{i}} \rightarrow - K_{\pm \underline{i}}$. However, this transformation would not preserve the $\SO(6)$-invariant tensor $\epsilon_{(6)}$, preventing us from easily reconstructing the generalised frame. By simply following the recipe we obtain\footnote{For completeness, we note that some signs and numerical coefficients in the formulas of \cite{Rovere:2025jks} have been corrected here. These corrections do not affect the results or conclusions of that work.}
\begin{equation}
\label{eq:sectionFinalIIB}
\begin{array}{rlll}
    &K_{i+} = &k_{i} &+ s\,d(2h_2^D Y^i)^+\\
    &&&- 
 d\left(\,(2h_1 h_2 + 2 h_1^Dh_2^D - \lambda) \,\text{vol}_2\,Y^i \right)  \\
 &&&+s  \smallstar d\left(\epsilon_{ijk} Y^j dY^k \wedge \text{vol}_2 \wedge\left(-\Lambda_-  dh_2^D + 4 h_2 h_2^D dh_1 \right)\right)^-\\ &&&+0\,,\\[2mm]
    &K_{\underline{i}+} = &0 &+s \,d(-2h_2 Y^{\underline{i}})^+\\
    &&&+ d\left(2\epsilon_{\underline{ijk}} Y^{\underline{j}} dY^{\underline{k}} \wedge (h_2 dh_1^D- h_1^D dh_2) \right) \\
    &&&+s  \smallstar d\left(\epsilon_{\underline{ijk}} Y^{\underline{j}} dY^{\underline{k}} \wedge \text{vol}_1 \wedge\left(\Lambda_-  dh_2 + 4 h_2 h_2^D dh_1^D \right)\right)^-\\
    &&& + 8\sqrt{g_{S^2 \times S^2}}(2h_1h_2 + 2 h_1^D h_2^D - \lambda)w h_1 h_2  k_{\underline{i}}^*  ,\\[3mm]
   &K_{i-} = &0 &+ d(-2 h_1 Y^i)^- \\
   &&&+s\,d\left(2\,\epsilon_{ijk} Y^{j} dY^{k} \wedge (h_1 dh_2^D - h_2^D dh_1)\right)\\
   &&&+ \smallstar d\left(\epsilon_{ijk} Y^j dY^k \wedge \text{vol}_2 \wedge\left(\Lambda_+ dh_1 + 4 h_1h_1^D dh_2^D \right)\right)^+\\
    &&& -s\, 8\sqrt{g_{S^2 \times S^2}}(2h_1h_2 + 2 h_1^D h_2^D + \lambda) w h_1 h_2  k_i^*,
   \\[2mm]
   & K_{\underline{i}-} = &sk_{\underline{i}} &+ d(-2 h_1^D Y^{\underline{i}})^- \\
   &&&+s\,d\left( \,(2h_1 h_2 + 2 h_1^Dh_2^D + \lambda) \,\text{vol}_1\,Y^{\underline{i}} \right)\\
   &&&+ \smallstar d\left(\epsilon_{\underline{ijk}} Y^{\underline{j}} dY^{\underline{k}} \wedge \text{vol}_1 \wedge\left(\Lambda_+  dh_1^D - 4 h_1 h_1^D dh_2 \right)\right)^+\\ &&&+0\,,
\end{array}
\end{equation}
where
\begin{equation}
    \Lambda_\pm = 2 h_1 h_2 + 2 h_1^D h_2^D \pm \lambda\,\hspace{5mm}\text{and}\hspace{5mm} w = \frac{1}{2} (h_1 h_2)^{-1}\star_\Sigma (dh_1 \wedge \star_\Sigma dh_2),
\end{equation}
and where $k_i^* = g(\cdot,\,k_i)$ and $k_{\underline{i}}^* = g(\cdot,\,k_{\underline{i}})$ with $g$ the round metric on $S^2 \times S^2$.

We remark that the sign $s$, specifying the gauging, enters the sections in a specific pattern. Splitting the sections in their various $p$-form components, the sign $s$ appears as
\begin{equation}
    \begin{array}{rccccccccl}
         & T\Mint & \oplus & T^*\Mint_{(+1)} & \oplus & T^*\Mint\otimes S&\oplus & T\Mint_{(+1)}\otimes S^* & \oplus & \Lambda^3 T^* \Mint\\
         K_{i+} =&   k_{i} & + & s b_i& + & s \tilde{b}_i & + & 0 & + & \lambda_i\\
         K_{\underline{i}+} =&   0 & + & s b_{\underline{i}}& + & s \tilde{b}_{\underline{i}} & + & \tilde{k}_{\underline{i}} & + & \lambda_{\underline{i}}\\
         K_{i-} =&  0 & + & b_i& + & \tilde{b}_i & + & s\tilde{k}_i & + & s\lambda_i\\
         K_{{\underline{i}}-} =&   sk_{{\underline{i}}} & + & b_{\underline{i}}& + & \tilde{b}_{\underline{i}}& + & 0 & + & s\lambda_{\underline{i}}\\
    \end{array}
\end{equation}
which sends the pairing $K_{Ia} \Omega K_{Jb}\rightarrow s K_{Ia}\Omega K_{jb}$. Since one of the compatibility constraints with respect to the $\SU(4)_S$ structure \cite{Malek:2017njj} requires
\begin{equation}
    K_{Ia} \Omega K_{Jb} \overset{!}{=} - 6 \kappa^2 \delta_{IJ} \epsilon_{ab} =  s \,32 \,\sin\theta_1\,\sin\theta_2\,(h_1h_2)^2 w
\end{equation}
we recover one of the constraints from \eqref{eq:posDefCondition}:
\begin{equation}
     - s (h_1h_2)\, \star_\Sigma (dh_1 \wedge \star_\Sigma dh_2) > 0\,.
\end{equation}

We have checked that all the compatibility constraints with respect to the generalised $\SU(4)_S$ structures as well as the torsion constraints are satisfied with this choice of generalised sections.

\section{Uplift formulas in type IIB}

We now translate the generalised-frame solution obtained in Section 3 into conventional type IIB supergravity fields. We recall that the Kaluza-Klein Ansatz is better written in terms of the ``KK-flat" fields according to the definition
\begin{equation}
    C_{(p)} = e^{\iota_{A_{KK}}} \bar{C}_{(p)}\,.
\end{equation}
In coordinates, this is equivalent to the familiar replacement
\begin{equation}
    dy^m \rightarrow Dy^m = dy^m + A^A \Theta_A{}^\alpha k_\alpha{}^m
\end{equation}
for any of the internal coordinates $y^m$ (e.g. $\bar{\Phi}_m Dy^m = \Phi_m dy^m$ for any one-form $\Phi_m$). From our ExFT ansatz, the KK vectors read 
\begin{equation}
    A_{KK} = A^A(x^\mu) \Theta_A{}^\alpha k_\alpha = A^{+i} k_i + s A^{-\underline{i}} k_{\underline{i}}\,.
\end{equation}
We reuse the notations of \cite{Rovere:2025jks}\footnote{To avoid confusion we only use the real coordinates on $\Sigma$ however the quantities $\scalP{a}{b}$ and $a \wedge b$ have a coordinate-free definition as $(1,\,1)$ forms on $\Sigma$.}
\begin{equation}
\begin{array}{rl}
    \scalP{a}{b} =& \frac{1}{2}\left(\partial_\eta a \partial_\eta a + \partial_\alpha b \partial_\alpha b\right),\\[2mm]
    a\wedge b =& \frac{1}{2}\left( \partial_\alpha a \partial_\eta b - \partial_\eta a \partial_\alpha b\right)\,,
\end{array}
\end{equation}
and define, based on a rescaling of the definitions in \cite{Assel:2011xz}, the quantities
\begin{equation}
    \begin{array}{rl}
    w =&  \scalP{\log h_1}{\log h_2}\,, \\
    n_1 =&  s\frac{\scalP{\log{h_1}}{\log{h_1}}}{\scalP{\log h_1}{\log h_2}}- e^\xi |\tau|^2\,,\\
    n_2 =&  s\frac{\scalP{\log{h_2}}{\log{h_2}}}{\scalP{\log h_1}{\log h_2}}- e^\xi\,,\\
    n_0 =& s\frac{\scalP{\log{h_1}}{\log{h_1}}}{\scalP{\log h_1}{\log h_2}}-e^{-\xi}\,.
    \end{array}
\end{equation}
Given \eqref{eq:posDefCondition}, the $n_i$ are all negative functions whereas $w$ must be negative for $s=1$ (AdS gauging) and positive for $s=-1$ (dS gauging).

The type IIB axio-dilaton fields are
\begin{equation}
\begin{array}{rl}
    \text{exp}(\Phi) =& -e^\xi\,\frac{h_1 }{h_2}\frac{n_0}{\sqrt{n_1  n_2}}\,,\\[2mm]
    C_0 =& \chi\frac{\,h_1\wedge h_2}{h_1^2\, n_0\, w}\,.
\end{array}
\end{equation}
The KK-flat metric reads
\begin{equation}
    \bar{ds}^2 = \Delta^{-1} \left(ds^2_{\text{ext}} - n_1^{-1} ds^2_{S^2_1}- n_2^{-1} ds^2_{S^2_2} - 2 s  w \,(d\alpha^2+d\eta^2) \right)
\end{equation}
and the warping factor is
\begin{equation}
    \Delta^{-4} = 16 n_1 n_2 h_1^2 h_2^2\,.
\end{equation}

The two-forms read
\begin{equation}
\begin{split}
    \bar{B}_2 =& -2 \chi e^\xi \frac{h_1}{n_1}\text{vol}_{S^2_1} + 2 s h_1^D \text{vol}_{S_2^2} + 2 h_1 \frac{\log h_1 \wedge \log h_2}{n_2 w}\text{vol}_{S_2^2}\,, \\
    &+A^{i-} d(2h_1 Y^i) + A^{\underline{i}-}d(2h_1^D Y^{\underline{i}})\,\\
    &+\left(B^{ij} + \frac{1}{2} A^{i+} A^{i-}\right)\epsilon_{ijk} (2 h_1 Y^k) + \frac{1}{2} A^{\underline{i}-}A^{\underline{i}+}s\, \epsilon_{\underline{ijk}} (2 h_1^D Y^{\underline{k}})\,,\\
    \bar{C}_2 =& 2 \chi e^\xi \frac{h_2}{n_2}\text{vol}_{S^2_2} + 2 s h_2^D \text{vol}_{S_1^2} - 2 h_2 \frac{\log h_1 \wedge \log h_2}{n_1 w}\text{vol}_{S_1^2}\\
    &+ A^{i+} s\, d(2h_2^D Y^i) + A^{\underline{i}+} s d(-2h_2 Y^{\underline{i}})\\
    &+ A^{i+} A^{j+} \epsilon_{ijk} (2h_2^D Y^i) + \left(B^{\underline{ij}} + \frac{1}{2} A^{\underline{i}-} A^{\underline{j}+} \right)s\,\epsilon_{\underline{ijk}} (-2 h_2 Y^{\underline{i}})\,.
\end{split}
\end{equation}
Finally the self-dual five form reads
\begin{equation}
\begin{split}
    \bar{\tilde{F}}_5 =&  \frac{1}{n_1 n_2}\text{vol}_{S^2\times S^2} \wedge\left[\star_\Sigma dj+ 4s \left( (1-e^\xi) h_2 dh_1 - (1-e^\xi|\tau|^2) h_1 dh_2\right)\right] \\
    &+\text{vol}_{\text{ext}} \wedge \left[ -dj + 4s \left( (1-e^\xi) h_2 dh_1^D - (1-e^\xi|\tau|^2) h_1 dh_2^D\right)\right] \\
    &+ 4 \frac{h_1 h_2}{n_1 n_2} \left(d\xi - e^{2\xi} \chi d\chi\right) \wedge\text{vol}_{1}\wedge\text{vol}_{2}\\
    &+ 8 h_1 h_2 w \star_{ext}\left(d\xi - e^{2\xi} \chi d\chi\right)\wedge d\alpha \wedge d\eta\\ 
    &-4(1+\bar{\star}) \Big(\mathcal{H}^{i+} (s Y^i h_1 dh_2\wedge \text{vol}_2\,n_2^{-1})\\
    &\phantom{-4(1+\bar{\star})\Big(}+ \mathcal{H}^{\underline{i}+} (s Y^{\underline{i}} h_1 dh_2^D\wedge \text{vol}_2\,n_2^{-1} + 2 h_1 h_2 w \,\epsilon_{\underline{ijk}} Y^{\underline{j}} dY^{\underline{k}} \wedge d\alpha \wedge d\eta)\\
    &\phantom{-4(1+\bar{\star})\Big(} +\mathcal{H}^{i-} ( Y^{i} h_2 dh_1^D\wedge \text{vol}_1\,n_1^{-1} + 2s h_1 h_2 w \,\epsilon_{ijk} Y^{j} dY^{k} \wedge d\alpha \wedge d\eta)\\
    &\phantom{-4(1+\bar{\star})\Big(} +\mathcal{H}^{\underline{i}+} (- Y^{\underline{i}} h_2 dh_1\wedge \text{vol}_1\,n_1^{-1})\Big)\\
    &+ \mathcal{H}^{MN} X_{MN}{}^P \left(\cdots \right)
\end{split}
\end{equation}
where
\begin{equation}
    dj = 4 \left[-(2+s)\star_\Sigma(h_1 dh_2 - h_2 dh_1) + \,d\left(\frac{h_1\wedge h_2}{w}\right)\right]\,.
\end{equation}
Due to the harmonicity of $h_{1,\,2}$, the r.h.s. is closed hence our definition of $j$ is locally well-defined.\footnote{This is actually a rewriting of the definition (5.37) of \cite{Rovere:2025jks} for $s=1$.} With this ansatz, $\tilde{F}_5$ is self-dual on-shell.

\section{An example of de Sitter S-fold}

The considerations of the previous sections provide a family of dS$_4$ solutions in type IIB obtained by fixing $\tau= i$ and setting all vectors and two-forms to zero. Such solutions should obey the Maldacena-Nunez no-go theorem \cite{Maldacena:2000mw}. However, we will see that for specific choices of harmonic functions $h_1$ and $h_2$, it is possible to perform a non-geometric quotient of the internal space known as ``S-folding". This will provide us with a compact, albeit non-geometric, de Sitter solution in type IIB. 

\subsection{The de Sitter S-fold solution}

The AdS$_4$ S-fold solution admitted a flat metric on $\Sigma$ and was obtained by choosing harmonic functions $h_1 = e^{\eta}\cos\alpha$ and $h_2 = e^{-\eta} \sin\alpha$. However, given the uplift ansatz, the metric for such harmonic functions would not be positive definite. Instead, we choose the harmonic functions
\begin{equation}
\begin{split}
    h_1 = e^\eta \sin\alpha\hspace{5mm} h_1^D = - e^\eta \cos\alpha\\
    h_2 = e^{-\eta} \sin\alpha\hspace{5mm} h_2^D = e^{-\eta} \cos\alpha
\end{split}
\end{equation}
With these choices of harmonic functions the condition \eqref{eq:posDefCondition} imposes that $\cot^2(\alpha) >1$, i.e., $\alpha \in [0,\,\pi/4]$. As such, it will prove convenient to abandon the conformal gauge on $\Sigma$ and we will perform the change of coordinates
\begin{equation}
\label{eq:covralpha}
    \cosh(r) = \cot(\alpha)\,.
\end{equation}
With these coordinates the metric reads:
\begin{equation}
    ds^2 = \Delta^{-1}\left[ds^2_{\text{dS}_4} + \frac{1}{2} \tanh(r)^2 \left(ds^2_{S^2_1}+ds^2_{S^2_2}\right) + \sinh^2r\, d\eta^2 + \frac{\sinh^4r}{(1+\cosh^2 r)^2} dr^2\right]
\end{equation}
where $ds^2_{\text{dS}_4}$ is the de Sitter external space with the cosmological constant fixed by the extermum of the scalar potential $\Lambda = 1$. The warping factor $\Delta$ reads
\begin{equation}
    \Delta^{-1} = \frac{2 \sqrt{2} \coth r}{\sqrt{1+\cosh^2r}}\,.
\end{equation}
The axio-dilaton is
\begin{equation}
    \tau = C_0 + i e^{- \Phi} = i e^{-2\eta}\,.
\end{equation}
The two-forms are
\begin{equation}
\begin{split}
    &B_2 = 2 e^\eta \sqrt{1+ (\cosh r)^{-2}}\, \text{vol}_{S_2^2}\,,\\
    &C_2 = -2 e^{-\eta} \sqrt{1+(\cosh r)^{-2}}\, \text{vol}_{S_1^2}\,.
\end{split}
\end{equation}
Finally the self-dual improved five-form reads
\begin{equation}
    \tilde{F}_5 = \frac{16 \cosh^2 r}{\sinh^3r\, (1+\cosh^2 r)}\, \text{vol}_{\text{dS}_4} \wedge dr + \frac{4}{\cosh^2 r}\text{vol}_{S_1^2}\wedge \text{vol}_{S_2^2}\wedge  d\eta\,.
\end{equation}

\paragraph{S-folding} We now observe that the solution can be quotiented along the $\eta$ direction upon introducing an $\SL(2,\,\mathbb{R})$ monodromy. Indeed, all the fields are independent of the $\eta$ coordinate except for the two-forms and the dilaton, both transforming non-trivially under $\SL(2,\,\mathbb{R})$. As such, the solution is invariant under $\eta \rightarrow \eta + T$ as long as it is compensated by the $\SL(2,\,\mathbb{R})$ duality transformation
\begin{equation}
    \mathfrak{M} = \begin{pmatrix} e^{T} & 0\\0 & e^{-T}
    \end{pmatrix}\,.
\end{equation}
Of course, such an element is not in the $\SL(2,\,\mathbb{Z})$ quantum duality group. As explained in \cite{Inverso:2016eet}, this can be cured by performing a global $\SL(2,\,\mathbb{R})$ transformation on the classical solution bringing the monodromy matrix to the form
\begin{equation}
    \mathfrak{M} = \begin{pmatrix}
        n & -1\\
        1 & 0
    \end{pmatrix} \hspace{1cm}n\in \mathbb{Z}_{\geq 3}\,,
\end{equation}
requiring that $T =  \log \tfrac{1}{2}(n + \sqrt{n^2-4})$. This procedure quantises the possible periodicities of the internal circle and bounds them from below. As such the internal circle $\mathbb{R}/(T\mathbb{Z})$ cannot be taken to be arbitrarily small.

\subsection{Compact internal space and effective Newton constant}

\paragraph{Compactness of the internal space}

First we study the volume of this manifold. We obtain that
\begin{equation}
    \text{Vol}(M_\text{tot}) = \text{Vol}_{\text{dS}_4} \times T \times (4\pi)^2 \times \underbrace{\int dr \Delta^{-5} \left(\frac{\tanh r}{2}\right)^4 \sinh r \frac{\sinh^2 r}{1+\cosh^2 r} }_{\mathcal{I}} \,.
\end{equation}
The factors of $4 \pi$ correspond to the area of the two-spheres and $T$ is the periodicity of the $\eta$ direction. We evaluate the last integral which is finite:
\begin{equation}
    \mathcal{I} = \int dr \frac{32 \sqrt{2} \cosh r \sinh^2 r}{(1+\cosh^2 r)^{7/2}}  = \frac{16 \sqrt{2}}{15}\,.
\end{equation}
This leaves us with an internal manifold of finite volume.

Although the original coordinate $\alpha$ takes values in a finite interval, the change of variables \eqref{eq:covralpha} maps this interval to the semi-infinite range \(r\in[0,\infty)\). We can nevertheless verify that the corresponding direction has finite proper length:
\begin{equation}
    L_r = \int dr \Delta^{-1/2} \frac{\sinh^2r}{(1+\cosh^2r)} \approx 2.7 < \infty\,.
\end{equation}
Taken together, these results show that the internal space, after the S-folding, is compact as previously stated. However, we do not obtain a proper scale-separated limit as the characteristic lengths in the internal space are of the same order of magnitude as that of the de Sitter radius. 
\paragraph{Effective Newton constant} Notwithstanding issues of scale separation, we compute the four-dimensional effective Newton constant
\begin{equation}
    \frac{1}{G_N^{(4)}} \cong \int d^dy \sqrt{\hat{g}}\, \Delta^{-4} = (4\pi)^2\, T \, \int dr 16 \frac{\sinh^3r}{(1+\cosh^2r)^3} = (4\pi)^2\,(4-\pi)\, T \,.
\end{equation}
This shows that although our construction evades the assumptions of the Maldacena-Nunez no-go, it is not in contradiction with it. Indeed, before performing the S-folding, the $\eta$ direction is not compact. This corresponds to the $T\rightarrow \infty$ limit implying a vanishing of the effective Newton constant $G_N^{(4)}$.

\paragraph{Regime of validity}
Finally, let us consider whether we stay in the regime of validity of two-derivative supergravity. Concerning $g_S$ corrections, we note that the dilaton profile is the same as that of the AdS S-folds. Since various holographic quantities have been checked in that regime \cite{Assel:2018vtq}, we can safely assume that $g_S$ corrections should not be a problem.

However, we show now that higher-derivative corrections are needed in order to realise this solution as a string-theory vacuum. This can be seen by computing the Ricci scalar of our solution
\begin{equation}
    R = \frac{\sqrt{2}}{128} \sqrt{1+\cosh^2 r} \frac{(118 + 111 \cosh 2r + 26 \cosh 4r + \cosh 6r)}{\cosh^3r \sinh^3r}\,.
\end{equation}
Since the Ricci scalar diverges at the boundaries of $\Sigma$, the dimensionless curvature invariants in string units cannot remain parametrically small throughout the internal space. Consequently, the two-derivative approximation necessarily breaks down sufficiently close to the boundary of $\Sigma$, i.e. when $r\rightarrow 0$ or $r\rightarrow \infty$. We conclude that we cannot safely ignore $\alpha'$ corrections when reaching the $r\rightarrow 0$ or $r\rightarrow \infty $ limits.

\section{Conclusion}

\subsection{Summary of results}

First, we classified the uplifts of the de Sitter gauging of $\SO(4)$-gauged $D=4$ $\mathcal{N}=4$ pure supergravity. Using generalised geometry, we have identified the corresponding $\SU(4)_S$ structure and solved the differential constraints determining the uplifts in type IIB. The uplift procedure provides solutions to type IIB supergravity on 
\begin{equation}
   M_{\text{tot}} =  \text{dS}_4 \times \Sigma \times S^2 \times S^2\,,
\end{equation}
with non-trivial warping and fluxes. The solutions are parametrised by a pair of harmonic functions $h_1$ and $h_2$ on $\Sigma$. These solutions are reminiscent of the familiar supersymmetric AdS$_4$ solutions of \cite{DHoker:2007hhe,DHoker:2007zhm}. Importantly, the positivity condition for the ten-dimensional metric differs by an overall sign from those in the AdS case. Consequently, the regions in which the respective solutions are well-defined do not overlap.

Second, we considered a particular member of this family for which the Riemann surface is a ribbon $\Sigma = \mathbb{R} \times I$. On this space, the internal geometry has a non-compact $\mathbb{R}$ direction and thus vanishing effective four-dimensional Newton constant. The fields are invariant under a translation along that direction only up to an $\SL(2,\,\mathbb{R})$ duality transformation. Taking the monodromy to lie in $\SL(2,\,\mathbb{Z})$, the translation can be combined with an $\SL(2,\,\mathbb{Z})$ transformation to define an S-fold quotient. The resulting solution has a finite internal volume and, in particular, a finite and non-vanishing effective Newton constant. This provides an explicit example in which a non-geometric identification evades the assumptions underlying the Maldacena-Nunez no-go theorem.

We stress that this solution has to be interpreted as a classical solution of the two-derivative type IIB equations of motion. Although the S-folding procedure has been well tested in a variety of holographic setups and does not seem to break the weak $g_s$ SUGRA approximation, we stress that this solution will require higher-derivative $\alpha'$ corrections to be interpreted as a proper string theory background. Moreover, this solution is not a scale-separated vacuum and suffers from perturbative instabilities.

\subsection{Outlook}

\paragraph{New de Sitter solutions}

This work illustrates how consistent truncations can be used as a solution-generating technique for de Sitter backgrounds and we hope that the same tool can be used to study more generic de Sitter U-folds with non-vanishing effective Newton constant. In particular, the explicit S-fold considered above relies on a particularly simple pair of harmonic functions, chosen such that the resulting solution is invariant under translations along the $\eta$ direction up to an $\SL(2,\,\mathbb{R})$ transformation. It would be interesting to determine whether more general choices of $h_1$ and $h_2$ can lead to qualitatively different global geometries and, in particular, to other de Sitter S-folds with finite effective Newton constant. A systematic analysis of the allowed global behaviour of the harmonic functions could also clarify whether the absence of scale separation found for the solution considered here is a generic feature of the dS family or merely a consequence of this particular choice. More generally, one could ask whether suitable choices of $\Sigma$, together with appropriate $\SL(2,\,\mathbb{Z})$ monodromies, can yield backgrounds with improved parametric control while retaining finite $G_N^{(4)}$.

We also note that the de Sitter S-fold admits a pair of ``flat-deformation" parameters $\chi_{1,\,2} \in\mathbb{R}/2\pi\mathbb{Z}$ as described for the AdS case in \cite{Guarino:2021kyp}. These two parameters break the $\SO(4)$ symmetry of the S-fold to a smaller $\SO(2) \times \SO(2)$ symmetry. Such deformations correspond to defining the solution on a mapping torus rather than on the usual direct product of the internal circle and the rest of the geometry. This immediately provides new de Sitter solutions to type IIB.

Finally, the same consistent truncation techniques should be applied to classify the M-theory uplifts of the three families of $\SO(4)$ gaugings. The uplifts of the anti-de Sitter gauging would provide new holographic backgrounds for 3d $\mathcal{N}=4$ SCFTs while the uplifts of the de Sitter gauging could provide new de Sitter solutions in M-theory. Moreover, a full classification of the Freedman-Schwarz gauging in type IIB is also missing. Such classifications go beyond the scope of this work.

\paragraph{Perturbative stability}

The solution presented here is unstable as was expected from \cite{Roest:2009tt}. Already at the level of the four-dimensional supergravity, gaugings admitting a perturbatively stable de Sitter vacuum are subjected to strong constraints \cite{Fre:2002pd}, and have been conjectured to be unobtainable for extended supergravity \cite{Andriot:2022way}. However, even when a stable de Sitter vacuum is found in a low-dimensional theory and uplifted to type II/11D supergravity, one would still have to study the perturbative stability of the truncated modes. If it is correct that stable de Sitter vacua cannot be embedded into a maximal lower-dimensional supergravity, even the KK-spectrometry techniques of \cite{Malek:2019eaz,Malek:2020yue} would not be applicable to compute the masses of all KK-modes around the solution. It would then be useful to  improve this method to the case of consistent truncations to non-maximal models. Nevertheless, we stress that considerations on the stability of the solution are not contained in the Maldacena-Nunez no-go and that non-geometric backgrounds do indeed provide a loophole to the no-go.

\paragraph{Complex type IIB and analytic continuations}

We saw that the de Sitter solutions are very closely related to the anti-de Sitter solutions of \cite{DHoker:2007hhe,DHoker:2007zhm}. However, the positivity condition on $h_1$ and $h_2$ is opposite in the AdS and dS cases. As such it is natural to wonder if these solutions can be interpreted as analytic continuations of one another or even if they can be more naturally understood as solutions in complexified type IIB along the lines of \cite{DHoker:2025nid}.

\paragraph{Symplectic frames and $\omega$-deformations}

We finally comment on a possible connection with the $\omega$-deformation of the $\mathcal{N}=8$ $\SO(8)$-gauged supergravity \cite{DallAgata:2012mfj}. In the embedding tensor formalism, the de Roo-Wagemans (dRW) angle, from which the three families of gaugings originate, has been interpreted as a choice of symplectic (or electric-magnetic) frame in the $\mathcal{N}=4$ theory \cite{Inverso:2015viq}. As such, the dRW angle has sometimes been referred to as the ``$\mathcal{N}=4$ analogue of the $\omega$-deformation". Given that the $\omega$-deformed theory does not admit an uplift to eleven-dimensional or type II supergravities \cite{Lee:2015xga,Inverso:2017lrz}, can our uplifts teach us something about a possible higher-dimensional origin of this deformation?

 At its core, the absence of an uplift is due to the non-existence of compatible solutions to the section constraint, namely solutions such that $\mathbb{P}\mathcal{E}^{\underline{m}}=\Theta_\omega{}^{\underline{m}}$ for $\omega \neq 0$. In the $\mathcal{N}=4$ case, we found that there exists an element $g\in\Es$ such that $\mathbb{P}\mathcal{E}=\Theta_{AdS}$ and $\mathbb{P}g\mathcal{E}=\Theta_{dS}$. We interpret this result as follows. The various dRW gaugings are inequivalent under the duality group $G_D^{\mathcal{N}=4}=\SL(2)\times \SO(6)$. However, when embedded into an ExFT framework, where the duality group is enlarged to $\Es$, the different symplectic frames can be related to one another by $\Es$ transformations. This is precisely the feature that is absent in the maximal supergravity case. The duality group $G_D^{\mathcal{N}=8}$ is already $\Es$ and there is then no $g_\omega\in\Es$ that can connect the different symplectic frames. Since it is not possible to weaken the section constraint of ExFT, this raises the question: is it possible instead to enlarge the $\Es$ duality group and construct a putative ``$\mathrm{U}(1)_\omega\times\Es$ ExFT" (or a larger generalised geometry such as constructed in \cite{Cederwall:2017fjm}) containing the $\omega$-deformed supergravity as a consistent truncation? We leave for future work whether such a theory exists, or whether it has any relevance to string theory.

\section*{Acknowledgements}
We thank Christoph Uhlemann for helpful discussions, as well as David Andriot, Thomas Van Riet, and Mario Trigiante for comments on the first version of this paper. C.S. is supported by a Postdoctoral Research Fellowship granted by the F.R.S.-FNRS (Belgium).

\appendix

\section{\texorpdfstring{$\Es$}{E7(7)}}
\label{app:Eseven}

Expressing the generators of $\es$ in terms of their $\mathfrak{sl}_8$ branching is particularly useful when discussing exceptional generalised geometry. However, it turns out that, since we are working with structure groups in $\SU(8)$, the $\su_8$ basis also has a role to play. We will collect here the various notations and definitions for $\es$ in both bases and explain how to go from one to the other.

\paragraph{SU(8) basis}

We start by defining the $\so_8$ $\gamma$-matrices:
\begin{equation}
\begin{array}{ll}
    \gamma_1 = \sigma_1 \otimes  \sigma_0 \otimes  \sigma_0 \otimes  \sigma_0\,,\hspace{5mm}
    &\gamma_2 = -\sigma_1 \otimes  \sigma_0 \otimes  \sigma_0 \otimes  \sigma_3\,,\\
    \gamma_3 = \sigma_1 \otimes  \sigma_1 \otimes  \sigma_2 \otimes  \sigma_2\,,
    &\gamma_4 = -\sigma_1 \otimes  \sigma_3 \otimes  \sigma_2 \otimes  \sigma_2\,,\\
    \gamma_5 = \sigma_2 \otimes  \sigma_2 \otimes  \sigma_3 \otimes  \sigma_0\,,
    &\gamma_6 = \sigma_2 \otimes  \sigma_2 \otimes  \sigma_1 \otimes  \sigma_0\,,\\
    \gamma_7 = \sigma_2 \otimes  \sigma_0 \otimes  \sigma_2 \otimes  \sigma_0\,,
    &\gamma_8 = -\sigma_1 \otimes  \sigma_2 \otimes  \sigma_0 \otimes  \sigma_2\,,\\
    \gamma_\star = \sigma_3 \otimes  \sigma_0 \otimes  \sigma_0 \otimes  \sigma_0\,,
\end{array}
\end{equation}
where the Pauli matrices are
\begin{equation}
    \sigma_0 = \mathbb{1}_{2\times 2} \,,\hspace{1cm} \sigma_1 = \begin{pmatrix}
        0 & 1\\
        1 & 0
    \end{pmatrix} \,,\hspace{1cm} \sigma_2 = \begin{pmatrix}
        0 & -i\\
        i & 0
    \end{pmatrix} \,,\hspace{1cm} \sigma_3 = \begin{pmatrix}
        1 & 0\\
        0 & -1
    \end{pmatrix} \,.
\end{equation}
This basis for $\gamma$-matrices is real and symmetric; they are block off-diagonal as this 16-dimensional representation splits into two 8-dimensional irreps according to the eigenvalues of the diagonal $\gamma_\star$ matrix, positive for $\mathbf{8}_s$ and negative for $\mathbf{8}_c$. The generators of $\mathfrak{so}_8$ in its three 8-dimensional representations are labelled by $\Gamma^v_{ab}$ for the 8$\times$8 antisymmetric matrix with the only non-vanishing entry at $ab$ equal to $1$ and 
\begin{equation}
\begin{split}
    \begin{pmatrix}
        \left(\Gamma^s_{[ab]}\right)_\alpha{}^\beta & 0\\
        0 & \left(\Gamma^c_{[ab]}\right)_{\dot{\alpha}}{}^{\dot{\beta}}
    \end{pmatrix} = \tfrac{1}{2} [\gamma_a ,\,\gamma_b]\,.
\end{split}
\end{equation}
Note also the (anti-)self duality properties of the rank-four gamma matrices:
\begin{equation}
\begin{split}
    &\Gamma^c_{[abcd]} = - \frac{1}{4!} \epsilon_{abcdefgh} \Gamma^{c}_{efgh}\\
    &\Gamma^s_{[abcd]} = \frac{1}{4!} \epsilon_{abcdefgh} \Gamma^{s}_{efgh}
\end{split}
\end{equation}
Which can be understood as the reducibility of the rank-four anti-symmetric representation of $\so_8$ into a self-dual and an anti-self-dual representation.

We now have all the prerequisites to write down the generators of $\es$ following the branching
\begin{equation}
\begin{array}{ccccc}
    \es &\rightarrow &\su_8 &\rightarrow &\so_8\\
    \mathbf{133} & \rightarrow & \mathbf{63} \oplus \mathbf{70} &\rightarrow &  \mathbf{28}_v \oplus \mathbf{35}_v \oplus \mathbf{35}_c \oplus \mathbf{35}_s\\
    \mathbf{56} &\rightarrow &\mathbf{28} \oplus \bar{\mathbf{28}}& \rightarrow &\mathbf{28}_v \oplus \mathbf{28}_v
\end{array}
\end{equation}
The $\mathbf{28}_v \oplus \mathbf{35}_v$ is the adjoint of $\su_8$. Given an $8\times 8$ traceless skew-hermitian matrix $u_a{}^b$ we write the corresponding $\es$ generator as 
\begin{equation}
    U_{M}{}^{N} = \begin{pmatrix}
       u_{[a}{}^{[b} \delta_{c]}{}^{d]} & 0\\
        & \bar{u}_{[a}{}^{[b} \delta_{c]}{}^{d]}
    \end{pmatrix}
\end{equation}
To avoid confusion with the generators in the $\SL(8)$ basis, these generators will be labelled $(t^v)_{(ab)} \in \mathbf{35}_v$ (which are purely imaginary, traceless and symmetric) and $t^{28}_{[ab]} \in \mathbf{28}$ (which are real and antisymmetric). The non-compact generators transforming in the $\mathbf{35}_c$ and $\mathbf{35}_s$ embed as
\begin{equation}
\begin{split}
    &(t^s{}_{\alpha}{}^\beta)_M{}^N=\begin{pmatrix}
        0 & \left(\Gamma^{s}_{[ab][cd]}\right)_\alpha{}^{\beta}\\
        \left(\Gamma^{s}_{[ab][cd]}\right)_\alpha{}^{\beta}& 0
    \end{pmatrix}\,,\\[2mm]
    &(t^c{}_{\dot\alpha}{}^{\dot\beta})_M{}^N=\begin{pmatrix}
        0 & \left(\Gamma^{c}_{[ab][cd]}\right)_{\dot\alpha}{}^{\dot\beta}\\
        -\left(\Gamma^{c}_{[ab][cd]}\right)_{\dot\alpha}{}^{\dot\beta}& 0
    \end{pmatrix}\,.
\end{split}
\end{equation}
In these notations, the pair of antisymmetric indices $[ab]$ labels an entry in the $\mathbf{28}$.

\paragraph{SL(8) basis}

We refer to \cite{Sterckx:2024vju} for the explicit form of the $\es$ generators $t_A{}^B$ and $t_{[ABCD]}$ in the $\SL(8)$ basis. We simply note that one can go from one basis to the other using the change of basis:
\begin{equation}
    S_{\su \rightarrow \sl} = \frac{1}{\sqrt{2}}\,\begin{pmatrix}
        \Gamma & 0\\
        0 & \Gamma 
    \end{pmatrix} \cdot \begin{pmatrix}\mathbb{1}_{28\times 28} & \mathbb{1}_{28\times 28} \\
   -i \mathbb{1}_{28\times 28}  & i\mathbb{1}_{28\times 28} \end{pmatrix}
\end{equation}
where $\Gamma = -\frac{1}{2} \text{Tr}(\Gamma^c_{[AB]} \cdot \Gamma^{v\,ab})\,.$
The two representations are related by
\begin{equation}
    \rho^{SU}(\cdot) = (S_{\su \rightarrow \sl})^\dagger \rho^{\SL}(\cdot) S_{\su \rightarrow \sl}\,.
\end{equation}

\bibliography{reference}

\end{document}